# The electron attachment to nitric oxide (NO) controversy

Ana I. Lozano,[1,2,3] Juan C. Oller,[4] Paulo Limão-Vieira,[2] and Gustavo García[1,*]

ABSTRACT:

We report novel total electron scattering cross sections (TCS) from nitric oxide (NO) in the impact energy range from 1 to 15 eV by using a magnetically confined electron transmission apparatus. The accuracy of the data to within 5% and its consistency across the energy range investigated, shows significant discrepancies from previous works as to the major resonance features and magnitude of the TCS. Within the shape of the TCS, we have identified nine features which have been assigned to electron attachment resonances, most of them reported for the first time, while a comprehensive analysis of those peaking at 7.0, 7.8 and 8.8 eV has led to solve the controversy about dissociative electron attachment (DEA) cross-section that persisted for more than 50 years.

1. INTRODUCTION

Nitric oxide (NO) is a very reactive molecule frequently interacting in biological and environmental media. In order to understand the underlying molecular mechanisms related to electron induced processes within these reactions, electron affinities of NO and resonant formation of the NO$^-$ anion (see Alle et al.[1] and references therein) have been experimentally analyzed. Also, calculations with the R-matrix method[2] and with the Schwinger multichannel method[3] evidenced the presence of resonances superimposed on the elastic scattering cross sections. These studies are mainly focused on the elastic and vibrationally excited resonances of the lowest-lying electronic state of the anion. In particular, Trevisan et al.[4] calculated the contribution of these resonances to the elastic and vibrationally inelastic cross section in the 0 – 2 eV energy range as well as the corresponding dissociative electron attachment (DEA) cross-section yielding O$^-$ ($^2$P) in the ground state. For higher energies, Rapp and Briglia[5] (1965) measured the total anion current generation from 6.5 to 13 eV, where a broad double peak structure attributed to DEA processes has been reported. Note that these are the only direct absolute negative ion formation cross-section values available in the literature to date. The previous mass spectrometric analysis of Hagstrum and Tate[6] (1941) concluded that O$^-$ was the only anion formed by DEA to NO. The onset potential obtained for the O$^-$ formation resonant process at 7.0 ± 0.3 eV, is higher than that obtained by Hanson (1937)[7] by extrapolating to zero the kinetic energy of the detected anions. Additional measurements from Chantry[8] (1968) provided the total anion current together with mass analysis and kinetic energy of the formed species. Again, only O$^-$ fragments were detected with an energy threshold of 7.5 ± 0.1 eV given by the linear fit of the O$^-$ kinetic energy distribution. Considering the available excited states of the neutral nitrogen atom, the energy thresholds ($E_{th}$) of the possible DEA processes are:

$e^-$+NO →[NO$^-$]*→O$^-$ ($^2$P) + N*($^4$S)     ($E_{th}$=5.07 eV)           (1a)

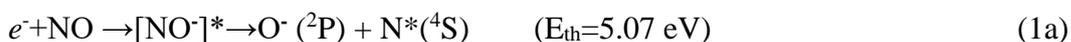

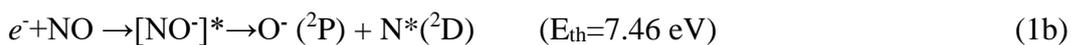
$$e^- + NO \rightarrow [NO^-]^* \rightarrow O^- (^2P) + N^* (^2D) \quad (E_{th}=7.46 \text{ eV}) \quad (1b)$$

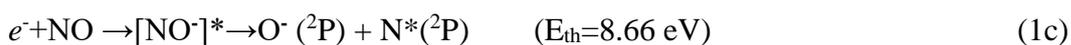
$$e^- + NO \rightarrow [NO^-]^* \rightarrow O^- (^2P) + N^* (^2P) \quad (E_{th}=8.66 \text{ eV}) \quad (1c)$$

with [NO$^-$]* the temporary negative ion (TNI). By comparing the energy thresholds calculated from the bond dissociation energy, electron affinity and the possible excitation energies of the remaining nitrogen atom with the corresponding experimental energy threshold, and considering the slope of the detected O$^-$ current as a function of the electron impact current (see Ref. 8 for details), Chantry concluded that the only process leading to N* in the $^2$D excited state (1b) can contribute to the observed O$^-$ fragment.

Further studies in this energy range provided relative or even absolute values of the dissociative attachment cross sections[9] through normalizing procedures or using the relative flow technique (see Ref. 9 and references therein). In spite of the experimental effort made, important discrepancies about the energy threshold and the intensity of the resonances leading to O$^-$ formation persisted. Some years later Orient and Chutjian[10], using combined electric and magnetic spectrometry techniques, concluded that the three channels (1a, 1b, 1c) contribute to the broad O$^-$ feature reported in Refs. 4 and 7, with reaction (1a) being clearly dominant and yet assigning cross section-values to all of them. This additional controversy motivated new studies based on more accurate techniques, using high resolution electron beams and time of flight mass spectrometry. Chu et al.[11] showed that Orient and Chutjian's conclusions were possibly wrong and stated that the dissociative attachment process to NO is dominated by reaction (1b) with a small contribution (15-20%) of reaction (1c). This conclusion was also supported by the mass spectrometry analysis of Denifl et al.[12] and later commented by Illenberger and Märk.[13] Nonetheless, Orient and Chutjian[14] replied to this comment suggesting that discrepancies can arise from the different angles of detection and claiming that their experimental system has the ability to collect ions ejected in all directions. From the persisted ambiguities within the scientific community, Allan[15] used a high-resolution spectroscopy technique to conclude that only signals corresponding to reaction (1b) yield a DEA process. In the conditions of Allan's experiment, without external magnetic fields, no signals arising from reactions (1a) and (1c) were detected. Thus, from the different rationales put forward, and after all the above discussions it seems that Chantry's interpretation was right and only one channel (1b) contributes to the anion generation by DEA to NO. Despite the considerable information, including higher energy Feshbach resonances[16], some doubts about the influence of specific experimental conditions on the results persist and relevant inconsistences in the absolute value of the DEA cross sections require further attention. More recently, Nandi et al.[17] combined accurate experimental techniques (viz. ion momentum imaging) with R-matrix scattering calculations with the aim of definitively clarifying the above controversy. From the experimental data they concluded that, in disagreement with Allan's electron spectroscopic analysis, O$^-$ detected signals can result from reactions (1b) and (1c).

In addition to previous reports on electron scattering by NO molecules[18,19], in 2019 Song et al.[20] reviewed the current situation of electron scattering cross sections from NO, N$_2$O and NO$_2$. As derived from this study, no direct electron attachment data are available in the literature. Only the aforementioned contradictory DEA results are reported. As shown by Alle et al.[1] for the lower energies, electron attachment (EA) resonances appear as relative maxima of the total cross-section (TCS) values. However, apart from those low-energy results[1], the TCS measurements reported by Song et al.[20] do not show any resonance signature above 2 eV (see Figure 2 of Ref. 16). These measurements were performed by Zecca et al.[21], Dalba et al.[22] and Szmytkowski and Maciag[23] and, considering the precedent discussion about electron attachment resonances leading to O$^-$ formation processes around 8 eV, a new inconsistency appeared in the electron-NO scattering literature. This contradiction was not discussed by Song et al.[20] and is probably due to deficiencies in the energy resolution used in these experiments, but it is clear that still requires experimental verification.

The above considerations, together with the recent interest in nitro-compounds as potential radiosensitizers[24,25,26] via NO and $NO_2$ generation and the relevance of these reactive species in the biosphere conditions[27] motivated the present study.

2. EXPERIMENTAL METHOD

Here we use a "state of the art" magnetically confined electron transmission apparatus to determine the total electron scattering cross sections from NO, for electron impact energies ranging from 1 to 15 eV, with a total uncertainty limit of about 5%. The considered energy range included all the expected electron attachment resonances. Details on the experimental setup and procedure can be found in a previous publication.[28] The quoted energy spread of the magnetically confined electron beam is better than 200 meV but the cut-off procedure followed to measure the transmitted beam intensity provided an effective energy resolution of about 50 meV (see Ref. 22 for details).

As already mentioned, if the energy resolution of the electron scattering experiment is good enough, resonances should appear as a clear increase of the TCS around the resonant energy.

3. RESULTS AND DISCUSSION

The results of the present TCS measurements, with total uncertainties within 5%, are shown in Figure 1. A detailed description of all the error sources and the analysis of the total uncertainty limits can be found in Ref. 28.

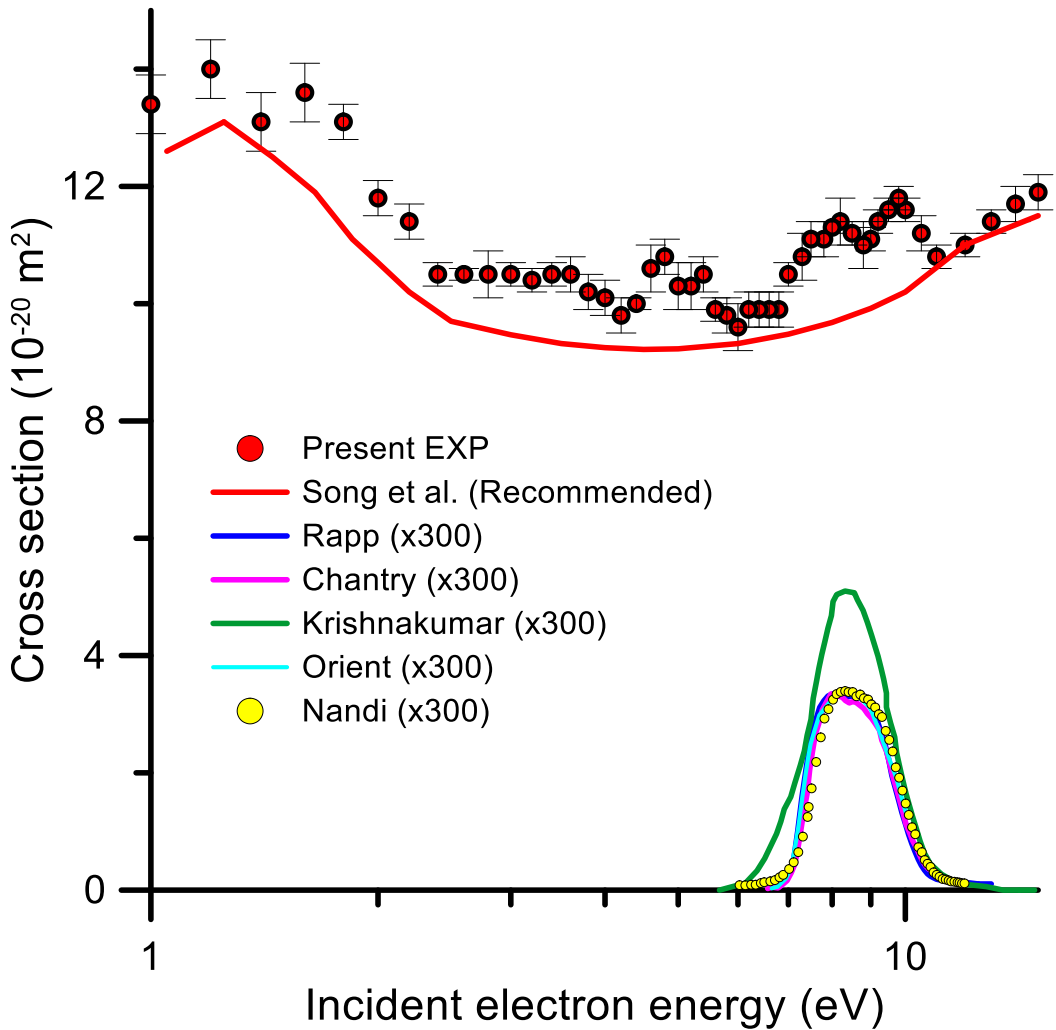

**Figure 1.** Total electron scattering cross sections (●, present experimental results; —, recommended values from Ref. 19) and O⁻ production by DEA cross sections

Figure 1 reveals that the TCS features we find between 3 and 10 eV are not accounted by the TCS values recently recommended by Song. et al.[20] As mentioned above these peaks on the TCS values correspond to the contribution of resonant EA processes. The TCS values corresponding to the minima of these features are in good agreement with the recommended data but these features were not detected in previous experiments. We have also plotted in this figure the O⁻ production by DEA cross sections discussed above. Such DEA resonances are within the prominent peaks we measured between 7 and 12 eV. Other relevant aspect is the absolute value assigned to the DEA cross-section is about two orders of magnitude lower than that corresponding to the EA peak.

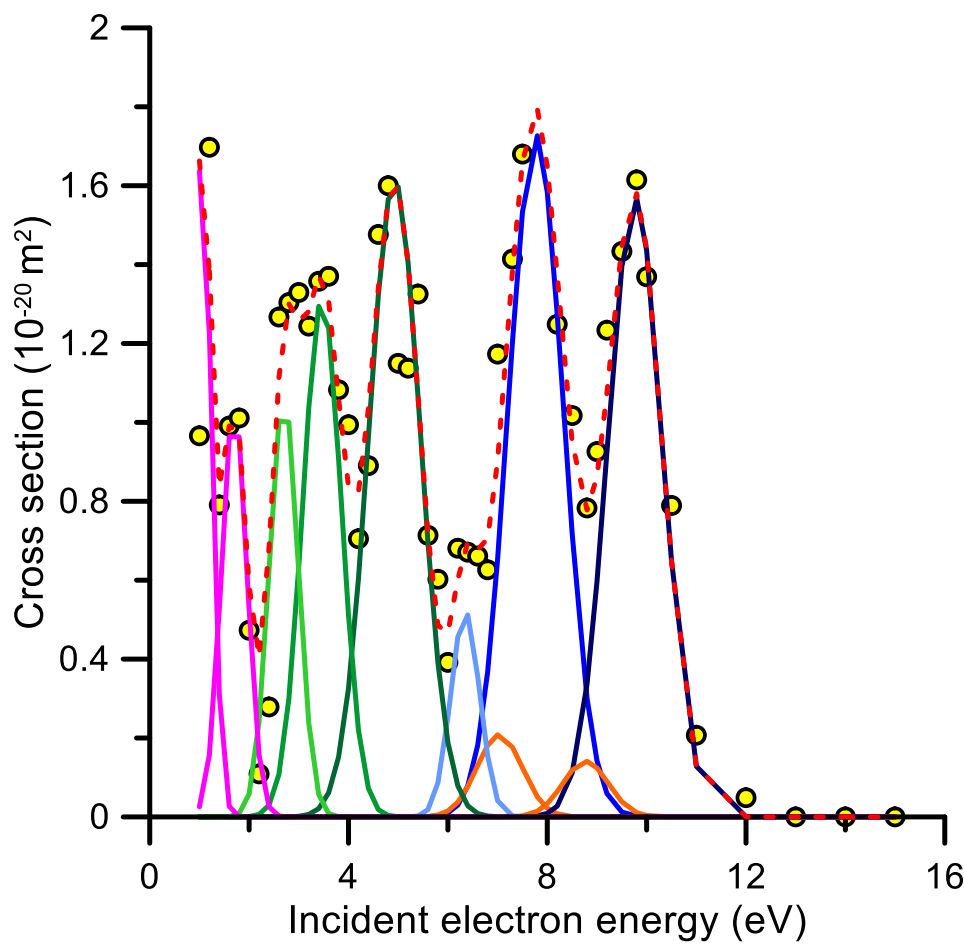

**Figure 2**. Electron attachment cross sections: ○, present results (see text for details); ---, Gaussian fit to the present experimental data.

For a deeper analysis of the experimental findings we have plotted in Figure 2 the EA cross-section values resulting from the subtraction of Song et al.[20] recommended elastic and electronic excitation cross sections to our TCS experimental values. As seen in this figure, a significant number of resonances appear between 1 and 12 eV, reaching peak values of about $1.7 \times 10^{-20}$ m$^2$. The intense peak at $7.8 \pm 0.2$ eV agrees in shape with that of O$^-$ production previously reported.[6,8] The threshold energy of this resonance is 7.5 eV, in agreement with that corresponding to reaction (1b), thus leading to O$^-$ formation and leaving the neutral nitrogen atom in the $^2$D state. However, its absolute value is 100 times higher than those reported in refs. 4 and 8. This may indicate that once formed the oxygen anion, most of the times the electron detaches from the atom faster than the experimental time of flight detection window. In both sides of this resonance feature we note two other prominent peaks with threshold energies of 5.1 and 8.7 eV. Note that these energies are very close to the threshold of reactions (1a) and (1c) DEA processes. These results evidence that O$^-$ anions are, in principle, formed via reactions (1a), (1b) and (1c) but the fast electron autodetachment associated with (1a) and (1c) reduces the probability of being detected within the detection timing of the different experimental systems.

Moreover, there are other channels compatible with the electron attachment to NO, i.e.:

$e^-+\text{NO} \rightarrow [\text{NO}^-]^* \rightarrow e^-+\text{NO}^*$    ($E_{th}$=<1 eV)     (1d)

$e^- + NO \rightarrow [NO^-]^* \rightarrow O(^3P) + N^-(^3P) \rightarrow e^- + O(^3P) + N$      ($E_{th}$=6.57 eV)      (1e)

$e^- + NO \rightarrow [NO^-]^* \rightarrow O(^3P) + N^-(^1D) \rightarrow e^- + O(^3P) + N$      ($E_{th}$=7.94 eV)      (1f)

$e^- + NO \rightarrow [NO^-]^* \rightarrow O(^1D) + N^-(^3P) \rightarrow e^- + O(^1D) + N$      ($E_{th}$=8.53 eV)      (1g)

Reaction (1d) is related to those resonances reported by Alle at al.[1], which were interpreted by Tennyson and Noble[2] and Trevisan et al.[3] (see these references for details). The resonances we find below 2 eV agree well with these, although our minimum energy limit (1 eV) does not allow to reproduce all the vibrational structure associated with such resonances. The weak resonances shown in Figure 2 at 7.0 and 8.8 eV, with threshold energies around 6 and 8 eV, can be associated to reaction (1e) and (1f). It is well-known that electrons autodetach from $N^-$ anions very fast and the negative nitrogen atom has never been detected in DEA experiments, while the (1g) process is not distinguishable within the present experimental conditions.

As can be seen in Figure 2, between 2 and 6 eV we found three peaks which have not been previously reported. The threshold energies of these resonances are much lower than those corresponding to the processes represented by (1e), (1f) and (1g) and, since no anion fragments have been detected in previous studies, we should assume that they arise from EA processes as those represented by reaction (1d).

## 4. CONCLUSIONS

In conclusion, using a "state-of-the-art" magnetically confined electron transmission apparatus we have measured the total electron scattering cross section for electron impact energies ranging from 1 to 15 eV. The present energy resolution (<50 meV) has allowed to detect different electron attachment resonances occurring in this energy range. We have identified the three EA processes leading to $O^-$ formation at the energy thresholds predicted by theory (5.0, 7.4 and 8.6 eV) for the corresponding states of the neutral N atom ($^4S$, $^2D$ and $^2P$, respectively). Since previous DEA measurement only detected $O^-$ formation from the $^2D$ channel (7.4 eV threshold energy), we have concluded that the electron autodetachment process is dominant in these transitions and the $O^-$ detection from the other two channels strongly depends on the experimental timing conditions. Transitions corresponding to $N^-$ formation have also been detected although from previous experimental evidence we have concluded that electron autodetachment is the only decay channel. Absolute EA cross sections have been derived by combining the present TCS measurements with the recommended integral elastic and electronic excitation cross sections. The present EA cross sections resulted to be two orders of magnitude higher than the available DEA cross sections, thus indicating that electron autodetachment is the most probable decay channel of the transient anion formed by electron attachment to NO. We consider that these findings contribute to resolve the dissociative electron attachment to NO controversy existing from 1965 and add accurate total electron scattering and electron attachment cross sections to the available databases which, as a consequence of this study, need to be updated. In order to characterize the new EA resonances reported in this experimental study, further accurate calculations including high energy electronic excited states of the NO$^-$ anion are needed.

AUTHOR INFORMATION

**Corresponding Author**

**Gustavo García** - *Instituto de Física Fundamental, Consejo Superior de Investigaciones Científicas, Serrano 113-bis, 28006 Madrid, Spain;* orcid.org/0000-0003-4033-4518; Email: g.garcia@csic.es

**Authors**

**Ana I. Lozano** - *Instituto de Física Fundamental, Consejo Superior de Investigaciones Científicas, Serrano 113-bis, 28006 Madrid, Spain; Laboratório de Colisões Atómicas e Moleculares, CEFITEC, Departamento de Física, Faculdade de Ciencias e Tecnologia, Universidade NOVA de Lisboa, Caparica 2829-516, Portugal; Institut de Recherche en Astrophysique et Planétologie (IRAP), Université Toulouse III - Paul Sabatier, 9 Avenue du Colonel Roche, Toulouse 31028, France;* orcid.org/0000-0003-4613-0372

**Juan C. Oller** - *División de Tecnología e Investigación Científica, Centro de Investigaciones Energéticas, Medioambientales y Tecnológicas, 28040 Madrid, Spain*

**Paulo Limão-Vieira** - *Laboratório de Colisões Atómicas e Moleculares, CEFITEC, Departamento de Física, Faculdade de Ciencias e Tecnologia, Universidade NOVA de Lisboa, Caparica 2829-516, Portugal;* orcid.org/0000-0003-2696-1152

**Notes**

The authors declare no competing financial interest.

ACKNOWLEDGEMENTS

This study has been partially supported by the Spanish Ministerio de Ciencia e Innovación (Project PID2019-104727RB-C21) and the European Association of National Metrology Institutes (Project 21GRD02 BIOSPHERE).